

\magnification=\magstep1  
\raggedbottom             
\tolerance=1000           
\baselineskip 24 truept   
\parindent 36 truept      
\hoffset .5 truein        
\hsize 5.6 truein         
\vsize 9.0 truein         

\centerline{\bf INFINITE VOLUME EXTRAPOLATIONS}

\centerline{\bf OF FINITE CLUSTER CALCULATIONS,}

\centerline{\bf ---HOW CORRECT ARE THESE?}
\bigskip
\bigskip
\centerline{S. Mazumdar, F. Guo, D. Guo, and K. -C. Ung}

\centerline{Department of Physics, University of Arizona}

\centerline{Tucson, AZ 85721, USA}
\medskip
\centerline{J. Tinka Gammel}

\centerline{Materials Research Branch, NCCOSC RDT$\&$E Division}

\centerline{San Diego, CA 92152-5000, USA}
\bigskip
\bigskip
\bigskip
\centerline{\bf ABSTRACT}

 Extrapolations of numerical data obtained from finite cluster calculations
to the infinite volume limit can often give incorrect results. We discuss
four separate cases: (a) the intensity of the lowest two-photon absorption
in the infinite polyene, (b) bond alternation in the infinite polyene, (c)
Cooper type pairing in the simple Hubbard model, and (d) pairing within the
extended Hubbard model.
\eject
\noindent {\bf I. Introduction}

Approximate analytic techniques have been known to lead to
qualitatively incorrect predictions for
solid state systems with strong electron correlation.
Exact numerical techniques are therefore often employed to
determine the behavior of such systems.
This approach consists of two steps, the first of which involves an
exact calculation for a finite cluster.
The second step involves an extrapolation of the finite
cluster data to the thermodynamic limit.
We point out that extrapolations are not
always valid, and in certain cases can lead to erroneous theoretical
predictions.

We believe that there is a strong necessity for such a discussion
currently. The technique of using exact numerical data for finite systems
to determine the behavior in the solid began with the classic calculations
for the Heisenberg chain by Bonner and Fisher$^1$, who, however, emphasized
the necessity of having one dimension and short range interaction.
In recent years, however, the numerical approach has been employed more
and more liberally, and calculations are routinely being done for higher
dimensions,
as well as for Hamiltonians with long range interactions.
Our hope is that the present
work will lead to more systematic studies of finite
size effects.

We discuss four specific cases here.
The correct
solutions to these problems are of interest to chemists and physicists.
These four different cases demonstrate different finite size effects,
and in each case
we show that literal interpretation of finite cluster calculations predict
{\it qualitatively} incorrect result.
\bigskip
\noindent {\bf II. Exact Finite Cluster Calculations}

\noindent {\bf A.} \underbar{The intensity of the two-photon absorption
(TPA) to the $2A_g$ and other}

\noindent \underbar{low
lying $A_g$ states in the infinite polyene}\ \ \ The occurrence of the
lowest two-photon state, the $2A_g$, below$^2$ the lowest one-photon
state, the $1B_u$ (note that optically inactive $B_u$ states need not be
considered), in finite polyenes necessarily requires a correlated
description for the $\pi$-electrons$^{2,3}$.
The entire discussion of TPA has, however,
so far been limited to the energy of the $2A_g$ state.
Here we extend this discussion to the question of the
intensity of the TPA to the $2A_g$ (and other low lying $A_g$
states) as a function of the chain length N.

 This discussion will be in the context of the extended Hubbard
Hamiltonian,
$$H=t\sum_{i,\sigma}[1-(-1)^i\delta](c_{i,\sigma}^+c_{i+1,\sigma}
+c_{i+1,\sigma}^+c_{i,\sigma})\ \ \ \ \ \ \ \ $$
$$+U\sum_in_{i,\uparrow}n_{i,\downarrow}
+V\sum_i (n_i-1)(n_{i+1}-1) \eqno(1) $$
\noindent where
all operators and
parameters have their usual meanings. We begin with the $U = V = 0$ limit.
In Fig.~1 we show the
occupancies of the highest two valence band levels and lowest two conduction
band levels for the $1A_g$ (the ground state), the $1B_u$, the $2A_g$ and the
$2B_u$ for arbitrary N.
TPA is given by
the third order nonlinearity $\chi^{(3)}(-\omega; \omega, \omega, -\omega)
\equiv \chi^{(3)}_{TPA}$,
$$\chi^{(3)}_{TPA}=\sum_{j,k,l}{\langle
1A_g|\mu|jB_u\rangle\langle jB_u|\mu|kA_g\rangle\langle kA_g|\mu|lB_u\rangle
\langle lB_u|\mu|1A_g\rangle\over (\omega_{jB_u}-\omega)
(\omega_{kA_g}-2\omega)(\omega_{lB_u}-\omega)}\ \  + \cdot \cdot \cdot
\eqno(2)$$
\noindent where $\mu$ is the dipole operator. Conventional wisdom
has been that TPA occurs every time an energy denominator goes to zero
in Eq.(2). We show below that this is not true.

 The intensity of the TPA to the $2A_g$ was calculated exactly for
$U = V = 0$ and for N = 6, 8, 10 and 12, typical chain lengths
for which calculations are done for nonzero U and V.
As seen in Fig.~2., TPA intensity increases with N here,
suggesting a large finite TPA
intensity for $N\to\infty$. This, however, is incorrect.

 The model in Eq. (1) has charge-conjugation symmetry, which implies
that for every single-particle valence band level at energy -$\epsilon$ there
exists a conduction band level at $\epsilon$. In the infinite chain, dipole
allowed absorptions occur only between these symmetrically placed molecular
orbitals, in agreement with $k$-conservation.
Thus for TPA to any intraband $A_g$ state, only two
``symmetric'' $B_u$ states need be considered. For the $2A_g$, these two $B_u$
states are the $1B_u$ and the $2B_u$. For any N, the absolute values of the
transition dipole moments $\langle 2A_g|\mu|1B_u\rangle$ and
$\langle 2A_g|\mu|2B_u\rangle$
are equal, while because of the fermion character of the electron
the products $\langle 1A_g|\mu|1B_u\rangle \langle 1B_u|\mu|2A_g\rangle$
and $\langle 1A_g|\mu|2B_u\rangle \langle 2B_u|\mu|2A_g\rangle$
are of opposite signs. At $N\to\infty$,
these two products become equal in magnitude, while having
opposite signs. This implies that the contribution to $\chi^{(3)}$
 in Eq.(2) by the
$2A_g$ vanishes$^4$, and the TPA intensity to the $2A_g$
and other band edge two-photon states is zero for $N\to\infty$.

We emphasize that the above analysis does not imply a zero intensity
for the total TPA to intraband two-photon states. Even with the
restriction to ``symmetric'' $B_u$ states, there can be very weak TPA
due to relatively high energy $A_g$ states that are slightly removed
from the band edge. For example, one can have such a high energy
intraband $A_g$ state that is coupled to the $1B_u$ and a $B_u$
state that requires excitation from  deep inside the valence band
to deep inside the conduction band. Because of the
small nonzero energy difference between the two $B_u$ states now, the
cancellation is no longer total. Weak TPA to states away from the band edge
is expected, and explains the shift of the TPA to higher energy from
the optical band edge in the long chain$^5$. Nevertheless, at infinite N,
TPA to band edge two-photon states goes to zero, in contradiction to the
prediction form Fig.~2. Since calculations of correlated chains are
limited to the region where TPA increases with N, the integration
of these two contradictory results is an important issue.

For the noninteracting case TPA can be calculated for arbitrary N. In
Fig.~3 we show the TPA intensity to $2A_g$
as a function of N. It is seen that the TPA intensity does go to
zero at very large N, but only after a maximum is reached.  Thus
calculations near $N = 10$ predict qualitatively incorrect results. On
the other hand, once the mechanism is understood, one can probe the
N-dependence of ${|S^-/S^+|}$, where $S^-$ and $S^+$ are the negative
and positive contributions to the TPA. This is shown in Fig.~4, where
the expected N-dependence is observed.

 We are now in a position to calculate TPA to $2A_g$ in long
correlated chains.  We have adopted the novel configuration
interaction (CI) approach of Srinivasan and Ramasesha$^6$ for
the interacting Hamiltonian of Eq. (1), with a cutoff 4t in the
energies of single-particle configurations. The approach can give
the $2A_g$ below the $1B_u$ even in long chains. In Table 1 we compare
calculated ${|S^-/S^+|}$ for many different U, V and $\delta$.
The ratio cannot be 1 in N = 20 - 30, but comparisons with the
noninteracting cases strongly suggest that the TPA to the $2A_g$
should be very weak in the long chain limit even for nonzero
Coulomb interaction, in contrast to what would be predicted from
exact evaluation$^7$ of the TPA near $N = 10$. This prediction is
in agreement with the tiny TPA to $2A_g$ in poly-BCMU polydiacetylene$^8$,
and the noted absence of the two-photon resonance in $\beta$-carotene$^9$.

\noindent {\bf B.} \underbar{The bond-alternation problem in the one
dimensional half-filled band}

\noindent Within the Su-Schrieffer-Heeger Hamiltonian,
$$H=\sum_{i,\sigma}[t-\alpha(y_i-y_{i+1})](c_{i,\sigma}^+c_{i+1,\sigma}
+c_{i+1,\sigma}^+c_{i,\sigma}) + {1\over 2}\kappa\sum_i(y_i-y_{i+1})^2
\eqno(3)$$
\noindent bond-alternation in the infinite polyene is unconditional.
Coulomb interactions can enhance this
bond alternation. For long range Coloumb interactions
$H_{ee}$ of the form ,
$$H_{ee}=U\sum_i n_{i,\uparrow}n_{i,\downarrow}
         +\sum_{i,j} V_j(n_i-1)(n_{i+j}-1)
\eqno(4) $$
\noindent two sharp inequalities give the ground state broken
symmetry$^{10}$. For off-site Coulomb
parameters that obey the convexity condition, $V_{j+1}+V_{j-1}>2V_j$,
the ground state can only be a bond-order wave (BOW) with alternating
bonds or a charge density wave (CDW) with
a periodic modulation of the on-site charge density.
The dominant broken symmetry is given by the inequality,
$$\sum V_{2n+1} \ \ \ \ \ \ \ \ {1\over 2}U+\sum V_{2n} \eqno(5) $$
\noindent where a smaller left hand side (LHS) gives BOW, while a smaller
right hand side (RHS) gives a CDW.
All Pariser-Parr-Pople (PPP) parametrizations   for the infinite polyene
then predict enhanced BOW$^{10}$.

The approach involving extrapolation technique was also employed in this
case, by Ramasesha and Soos$^{11}$. These authors calculated the electronic
energy gained for fixed bond alternation,
when compared to the equal bond length
case, for open even chains, closed $N = 4n$ rings,
and closed $N = 4n + 2$ rings,
and extrapolated the energy gains in the three series against $1/N$
for the PPP-Ohno Hamiltonian.
Enhanced energy gains compared to the simple H\"uckel case was taken
to be the proof of enhanced BOW.
While the result for the carbon-based polymers agrees with the
prediction from Eq. (5), let us examine this approach in detail.

The extrapolation of the 4n-ring series is invalid, since
the RHS of Eq. (5) in this case always has an
extra term that precludes the CDW entirely and
guarantees enhanced BOW.
The even chain extrapolation has
many different finite size effects. First, even chains have an odd number of
bonds, thus precluding equal bond lengths for small N. Second, the range of the
Coulomb interaction within PPP models is longer than the N for which
calculations are being done. This leads to a different Hamiltonian for each N.
Finally,
real space many-electron configurations
have entirely different matrix elements of $H_{ee}$ within the
open chain boundary condition and for the periodic ring. For example, the
configuration ...20202020..., where the numbers denote site-occupancies, has
a matrix element ${4U - 8V_1 + 8V_2 - 8V_3 + 4V_4}$ in the ring, but the
matrix element is ${4U - 7V_1 + 6V_2 - 5V_3 + 4V_4 - 3V_5
+2V_6 - V_7}$ in the open chain.
This last difference shifts the boundary between the BOW and the CDW
considerably, such that the open chain would predict enhanced BOW even as
Eq. (5) predicts a CDW. Since the $4n + 2$ ring series can only have a few data
points, this makes the extrapolation technique itself suspect.

 We demonstrate the above by doing the open chain calculation for the
PPP-Ohno parameters corresponding to polysilane. It has been assumed that
this has a simple BOW ground state with very large alternation$^{12}$.
We choose a U of 9.04 eV
and evaluate $V_j$ from the PPP-Ohno relationship,
$$V_j={U\over [1+0.39R_j^2]^{1/2}}
\eqno(6) $$
\noindent where the $R_j$ are the distances between the polysilane ``sites'',
and are taken from literature$^{12}$.
Eq. (5) predicts a CDW, with
different charge densities on the orbitals on the
same silicon atom.
The extrapolation technique nevertheless predicts enhanced
BOW, as seen in Fig.~5.

 Our motivation here was to demonstrate the inapplicability of the
extrapolation technique for the bond alternation problem and not to
claim that polysilane is a CDW. Whether or not polysilane is a CDW
depends on whether the PPP-Ohno model is truly appropriate for
polysilanes.  It is, however, interesting to note that the occurrence
of the lowest two-photon absorption considerably above the lowest
one-photon absorption$^{13}$ occurs naturally within the CDW model.

\noindent {\bf C.} \underbar{Pairing in Hubbard models}\ \ \
A great deal of numerical effort has gone into investigating
whether pairing of the Cooper type occurs in the weakly doped two-dimensional
Mott-Hubbard insulator. More recently, such a pairing mechanism has been
proposed$^{14}$
for doped $C_{60}$. The two dimensional calculations have now been
extended to the case of weakly coupled layers$^{15}$.
Specifically, these calculations involve determination of the ground
state energies of the half-filled band system, and the electronic energies
needed to add one and two electrons (holes). Pairing is supposed
to occur if the quantity
$$\Delta E=2E(N+1)-E(N)-E(N+2) \eqno(7)$$
\noindent is positive. In spite of considerable effort, it is still not
entirely clear whether pairing occurs in two dimension or not, or what
the actual mechanism of pairing is in the various cases.

We focus here on the mechanism of the pairing, and what this says
about pairing in the infinite system.
Fye et al$^{16}$ showed that pairing of the above kind is not limited
to two or three dimension, but can also occur in one
dimensional periodic rings with $N = 4n$ atoms.  The limitation
to 4n rings is very suggestive, as small 4n-rings undergo Jahn-Teller
distortions. In Fig.~6 we show our results of the pairing
calculations for N = 4 and 8.
The pairing disappears for the Jahn-Teller distorted rings.  To
understand this better, we calculated the effect of the Hubbard U on the
Jahn-Teller distortion of the 4n and $4n - 1$ electron systems.
These results are shown in Fig.~7. It is seen that U strongly reduces
the gain in energy for the Jahn-Teller distortion of the even (4n)
electron system, but the distortion of the odd numbered (4n -1)
electron system remains virtually unaffected. This
then gives us the mechanism of the pairing.
The information regarding the tendency to distort
is built into the wave function
in the form of the various bond order parameters.
By not letting the system distort, we are overestimating the energies
in each case, but as Fig.~7 indicates, this overestimation is much
stronger for the odd electron case than for the even electron case.
The calculated energy required
to add one hole (or electron) to the undistorted system is therefore
too high, thereby giving ``pairing''.

This same mechanism can explain ``pairing'' in
higher dimension. In Fig.~8 we show the results of our calculation for the
simple cube, and for various distorted forms.
For the undistorted simple cube, we merely reproduce the results
of White et al.$^{17}$, --- two distinct regions are found
in which pairing is seen to
occur. For the cube stretched along one axis,
we see that the pairing still exists. This is not the
Jahn-Teller mode for the cube, the noninteracting limit still
having degeneracies. However, the pairing at small U has disappeared
for both the Jahn-Teller distorted cube as well as the structure
with all three hopping integrals different. This clearly indicates
the relationship between pairing and the degeneracies in the
single particle limit. We have once again calculated the effect
of the correlation parameter U on the Jahn-Teller distortions.
The result is identical to that in Fig.~7, the tendency to distort
is almost unaffected for the 7 electron system, but for the
6 electron system this is strongly suppressed. The mechanism
of pairing at small U in the cube is then the same as in the 4n
ring. In the absence of distortion,
the energy to add one hole or electron is being overestimated.
\medskip
\noindent We have repeated such calculations for a number of three
dimensional molecules, and have found the following. For an N site
system, pairing at small U occurs only
if the $(N - 1)$  (or $(N + 1)$) electron system
has a strong tendency to have a Jahn-Teller distortion.
Thus the mechanism of pairing
at small U in all these case is related to the suppression of Jahn-Teller
distortion. Since the Jahn-Teller
distortion results from the discrete level degeneracies in
finite molecules,
we conclude that the observed pairing in
in the small U region in similar calculations is a finite size effect.

Fig.~8 explains the oscillation in the pairing energy of the
cube. Pairing at large U is unrelated to the tendency to have
Jahn-Teller distortion. We have occasionally found such pairing at
large U also in other systems, though not necessarily accompanied by pairing
at small U. The occurrence of such large U pairing in nonbipartite systems
(for e.g., in a pentagonal prism) indicates that this is also unrelated to
antiferromagnetism. Currently we are investigating whether the large U pairing
is also a finite size effect, and if so, the nature of the finite size effect.
The present work indicates the absence of pairing in the two dimensional
Hubbard model, as well as in weakly coupled layers.

\noindent {\bf D.} \underbar{Pairing due to intersite interactions}\ \ \
Many calculations exist that
find a pairing driven by intersite interactions rather than the on-site Coulomb
interaction. Since the intersite interactions in these
calculations are different, it is impossible to give a general critique.
Nevertheless, we believe numerical calculations here would in most cases find
phase segregation, rather than simple pairing, if adequate numbers of electrons
or holes are added to the half-filled band. In the case of the t-J model, this
tendency to phase segregate has been discussed$^{18}$. We choose one
simple example, a calculation claiming pairing driven by the nearest neighbor
Coulomb interaction$^{19}$ in the two-dimensional quarter-filled band, to
illustrate our point.

For large U and V, the quarter-filled band is a CDW.
The configuration with one extra
particle is as shown in Fig.~9. This has configuration interaction with a
higher energy configuration which is at energy 2V relative to it. If, however,
one adds two particles which occupy neighboring empty sites in the CDW, the
lowest energy configuration has configuration interaction with a configuration
that is at an energy V relative to it. Thus, CI should give pairing in two
dimension, though not in one dimension. This was actually numerically
demonstrated by Mazumdar and Ramasesha$^{19}$.
The fallacy in this argument becomes clear once three
particles are added to the system as shown in Fig.~9  Now, the lowest
energy configuration has CI with a configuration at the same energy, so that
a cluster of three particles is favored over a cluster of two and an isolated
particle. Similar qualitative arguments favor phase segregation in
related models. Thus, once again, thus, the pairing found within
such calculations is a finite size effect.
\bigskip
\noindent {\bf III. Conclusions}

The major conclusion of this presentation is that there are many different
finite size effects, and even when a given approach is satisfactory for
the determination of one property of a given class of materials (e.g., the
energy of the $2A_g$ in the infinite conjugated polyene), it does not
necessarily mean that the same approach is satisfactory for other properties
(in this case ground state bond alternation and the intensities of TPA).
Finite size effects should be investigated for each different problem
separately.  Probably the most useful approach to avoid the pitfalls is
to have a physical and intuitive mechanism of the effect in question in mind.
However, this can hardly be prescribed as a general approach. One simple way to
investigate finite size effects is to carefully investigate the size dependence
of the property in question within the noninteracting model, for which
calculations can be done for arbitrary sizes. Even though this procedure
is simple (and an obvious step), finite correlated clusters are often compared
to the infinite noninteracting system, simply because the latter is understood.
Such a procedure can, and often does, erroneously ascribe effects due to
finite size to correlations.

\eject
\noindent {\bf Acknowledgement}

Work at Arizona was supported by NSF grant No. ECS-89-11960. J. T. Gammel
was supported by a NRC-NRaD Research Associateship through grants
from the ONR and DOE.

\bigskip
\bigskip
\centerline{\bf REFERENCES}
\medskip
\item{1.} J. C. Bonner and M. E. Fisher,
{\it Phys. Rev.} {\bf 135}, A640, (1964).
\item{2.} B. S. Hudson, B. E. Kohler, and K. Schulten, {\it Excited States}
{\bf 6}, edited by E. C. Lim, (Academic, New York) (1982).
\item{3.} K. Schulten, I. Ohmine, and M. Karplus,
{\it J. Chem. Phys.} {\bf 64}, 4222, (1976).
\item{4.} Dandan Guo, S. Mazumdar, S. N. Dixit, {\it Nonlinear Optics}, (1993).
\item{5.} G. P. Agrawal, C. Cojan, and C. Flytzanis,
{\it Phys. Rev. B} {\bf 17}, 776, (1978).
\item{6.} B. Srinivasan and S. Ramasesha,
{\it Solid State Commun.} {\bf 81}, 831, (1992).
\item{7.} Z. G. Soos and S. Ramasesha,
{\it J. Chem. Phys.} {\bf 90}, 1067, (1989).
\item{8.} W. E. Torruellas, K. B. Rochford, R. Zanoni, S. Aramaki,
and G. I. Stegeman, {\it Opt. Commun.} {\bf 82}, 94, (1991).
\item{9.} J. B. van Beek, F. Kajzar, and A. C. Albrecht,
{\it J. Chem. Phys.} {\bf 95}, 6400, (1991).
\item{10.} S. Mazumdar and D. K. Campbell,
{\it Phys. Rev. Lett.} {\bf 55}, 2067, (1985).
\item{11} Z. G. Soos and S. Ramasesha,
{\it Phys. Rev. B} {\bf 29}, 5410, (1984).
\item{12.} R. G. Kepler and Z. G. Soos,
{\it Phys. Rev. B} {\bf 43}, 12530, (1991).
\item{13.} J. R. G. Thorne, Y. Ohsako, R. M. Hochstrasser, and J. M. Zeigler,
{\it Chem. Phys. Lett.} {\bf 162}, 455, (1989).
\item{14.} S. Chakravarty, M. P. Gelfand, and S. Kivelson,
{\it Science} {\bf 254}, 970, (1991).
\item{15.} D. G. Kanhere, in present proceedings.
\item{16.} R. M. Fye, M. J. Martins, and R. T. Scalettar,
{\it Phys. Rev. B} {\bf 42}, 6809, (1990).
\item{17.} S. R. White, S. Chakravarty, M. P. Gelfand, and S. A. Kivelson,
{\it Phys. Rev. B} {\bf 45}, 5062, (1992).
\item{18.} V. J. Emery, S. A. Kivelson, and H. Q. Lin,
{\it Phys. Rev. Lett.} {\bf 64}, 475, (1990).
\item{19.} S. Mazumdar and S. Ramasesha,
{\it Synth. Metals} {\bf 27}, A105, (1988).

\eject
\centerline{\bf Table}
\vskip2truecm

\noindent Table 1. The ratio of the negative and positive contributions to the
$2A_g$ TPA in long polyene chains within the extended Hubbard model.
\vskip1truecm
\baselineskip 12 truept           
$$\vbox{\settabs
\+~~~~~~$N$~~~~~&~~~~~$U$&~~~~~$V$&~~~~~$\delta$&~~~~~$|S^-/S^+|$~~~~~~~\cr
\hrule
\smallskip
\hrule
\+ ~~~$N$ &$U$ &$V$ &$~\delta$ & $|S^-/S^+|$ \cr
\smallskip
\hrule
\smallskip
\+ ~~~20  &0 &0 &0.3 &~~~0.79  \cr
\+ ~~~20  &3 &0 &0.3 &~~~0.80  \cr
\+ ~~~20  &6 &0 &0.3 &~~~0.78  \cr
\+ ~~~20  &4 &1 &0.3 &~~~0.81  \cr
\+ ~~~20  &0 &0 &0.1 &~~~0.21  \cr
\+ ~~~20  &6 &0 &0.1 &~~~0.67  \cr
\+ ~~~20  &4 &1 &0.1 &~~~0.60  \cr
\smallskip
\hrule
\smallskip
\+ ~~~24  &0 &0 &0.3 &~~~0.86  \cr
\+ ~~~24  &4 &0 &0.3 &~~~0.85  \cr
\+ ~~~24  &8 &0 &0.3 &~~~0.92  \cr
\+ ~~~24  &4 &1 &0.3 &~~~0.86  \cr
\smallskip
\hrule
\smallskip
\hrule}$$
\baselineskip 24 truept           

\eject
\bigskip
\centerline{\bf Figure Caption}

\noindent Figure 1. The single-particle orbital electron occupancies of the
$1A_g$, $1B_u$, $2A_g$ and the $2B_u$. Only the highest (lowest) two valence
(conduction) band levels are shown. All other valence (conduction) band levels
are filled (empty).
\medskip
\noindent Figure 2. The intensity of the $2A_g$ TPA in short finite chains
within the H\"uckel model. Here $E_g$ is the $1B_u-1A_g$ gap. Notice
that the intensity increases with $N$.
\medskip
\noindent Figure 3. The intensity of the $2A_g$ TPA as a function of $N$ within
the H\"uckel model. TPA goes to zero at $N\to \infty$.
\medskip
\noindent Figure 4. The ratio of the negative and positive contributions to the
$2A_g$ TPA as a function of $N$ within the H\"uckel model.
\medskip
\noindent Figure 5. The electronic energy difference per site between dimerized
and the equal bond length even chains for the parameters of Eq. (6). The
extropolated $N\to\infty$ values lies above the value for the dimerized
infinite H\"uckel ring.
\medskip
\noindent Figure 6. Pair binding energies (see Eq. (7)) for the (a) $N=4$ and
(b) $N=8$ rings. In each case the solid curve corresponds to the undistorted
ring, the dashed curve to the Jahn-Teller distorted ring.
\medskip
\noindent Figure 7. The gain in electronic energy on Jahn-Teller distortion
for the $N=4$ ring, as a function of Hubbard $U$. The solid and dashed curves
correspond to the 4 and 3 (5) electron cases respectively.
\medskip
\noindent Figure 8. Pair binding energies in the simple cube and three
distorted forms. (i) - - - - - -, simple cube; (ii) ---------, stretched
cube, $t_1=t_2>t_3$; (iii) --- --- ---, Jahn-Teller distorted cube;
(iv) --- - --- -,
all three hopping integrals different. Notice absence of pairing at
small $U$ for cases (iii) and (iv).
\medskip
\noindent Figure 9. Quarter-filled extended Hubbard CDW with (a) one extra
particle, (b) two extra particles, (c) three extra particles (crosses are
particles, dots empty sites). CT excitation in (a) costs energy $2V$, in (b)
$V$, and zero in (c).

\end